\documentclass{jaa}
\usepackage{natbib}
\usepackage{multicol}
\usepackage{caption}
\usepackage{graphicx}
\usepackage{multirow}
\usepackage{subfigure}
\usepackage{xcolor}
\usepackage{adjustbox}
\usepackage[colorlinks=true,linkcolor=blue,citecolor=blue]{hyperref}
\begin{document}

\title{Confirmation of interstellar phosphine towards asymptotic giant branch star IRC+10216}

\author{Arijit Manna\textsuperscript{1}, Sabyasachi Pal\textsuperscript{1,*}}
\affilOne{\textsuperscript{1}Department of Physics and Astronomy, Midnapore City College, Paschim Medinipur, West Bengal, India 721129\\}

\twocolumn[{
\maketitle
\corres{sabya.pal@gmail.com}


\vspace{0.5cm}
\begin{abstract}
Phosphorus (P) is an important element for the chemical evolution of galaxies and many kinds of biochemical reactions. Phosphorus is one of the crucial chemical compounds in the formation of life on our planet. In an interstellar medium, phosphine (PH$_{3}$) is a crucial biomolecule that plays a major role in understanding the chemistry of phosphorus-bearing molecules, particularly phosphorus nitride (PN) and phosphorus monoxide (PO), in the gas phase or interstellar grains. We present the first confirmed detection of phosphine (PH$_{3}$) in the asymptotic giant branch (AGB) carbon-rich star IRC+10216 using the Atacama Large Millimeter/Submillimeter Array (ALMA) band 6. We detect the $J$ = 1$_{0}$--0$_{0}$ rotational transition line of PH$_{3}$ with a signal-to-noise ratio (SNR) of $\geq$3.5$\sigma$. This is the first confirmed detection of phosphine (PH$_{3}$) in the ISM. Based on LTE spectral modelling, the column density of PH$_{3}$ is (3.15$\pm$0.20)$\times$10$^{15}$ cm$^{-2}$ at an excitation temperature of 52$\pm$5 K. The fractional abundance of PH$_{3}$ with respect to H$_{2}$ is (8.29$\pm$1.37)$\times$10$^{-8}$. We also discuss the possible formation pathways of PH$_{3}$ and we claim that PH$_{3}$ may be created via the hydrogenation of PH$_{2}$ on the grain surface of IRC+10216.
\end{abstract}

}]
\doinum{xyz/123}
\artcitid{\#\#\#\#}
\volnum{000}
\year{2021}
\pgrange{1--11}
\setcounter{page}{1}
\lp{11}

	\section{Introduction}
\label{sec:intro}
Phosphorus is one of the rare elements in the interstellar medium (ISM). Phosphorus is the 13$^{th}$ element in the meteoritic material and the 11$^{th}$ element in the crust of Earth \citep{mac97}. In ISM, the prebiotic chemistry of phosphorus (P) has attracted attention in astrochemical communities because P-bearing molecules, such as phosphorus monoxide (PO) and phosphorus nitride (PN), play important roles in the production of phospholipids and nucleic acids \citep{tur90, fon16}. Phospholipids and nucleic acids are important for the formation of life on our planet. P-bearing molecules may play an important role in the production of large complex biomolecules that store genetic information in nucleic acids \citep{mac97}. P-bearing molecules also play major roles in the synthesis of DNA and RNA \citep{mac97}. Earlier millimeter and submillimeter wavelength observations indicated the depletion of P-bearing molecules by a factor of $\geq$100 with respect to the cosmic abundance of P ($\sim$3$\times$10$^{-7}$) in cold and dense parts of the ISM \citep{tur90, fon16, lef16}. The depletion of P-bearing molecules suggests that the bulk of P-bearing species is blocked by interstellar grains \citep{fon16}. Recently, the ESA Probe Rosetta and the ALMA demonstrated that P-bearing species came to Earth through comets \citep{al16, riv20}. Comets carry several biomolecules because they travel between several star-forming regions \citep{al16}. In ISM, P-bearing molecules, such as PO, PN, HCP, CCP, and CP, were detected in the envelopes around evolved stars \citep{gu90, ten07, hal08, mil08, ten08, de13, zi18}. 

\begin{table*}{}
\centering
\scriptsize
\caption{Stellar properties of IRC+10216.}
\begin{adjustbox}{width=0.95\textwidth}
\begin{tabular}{cccccc}
\hline 
Physical properties&Value&Ref\\
\hline
$\alpha$(J2000)&09:47:57.4 &SIMBAD data base\\
$\delta$(J2000)&+13:16:43.5&SIMBAD data base\\
Distance&130 pc&\citet{men12}\\
Pulsation period&639$\pm$4 days&\citet{gro12}\\
$T_{\rm eff}$&$\sim$2200 K& \cite{men12}\\
Photospheric diameter (optical)&3.8 AU&\citet{men12}\\
Luminosity&8640$\pm$430 \(\textup{L}_\odot\)&\citet{men12}\\
Initial mass&3--5 \(\textup{M}_\odot\)&\citet{gue95}\\
Current mass&0.7--0.9 \(\textup{M}_\odot\)&\citet{la10}\\
Mass-loss&2$\times$10$^{-5}$ \(\textup{M}_\odot\) yr$^{-1}$&\citet{cr97}\\
$V_{\rm outflow}$&14.6$\pm$0.3 km s$^{-1}$&\cite{kn98}\\
$V_{\rm LSR}$&--26$\pm$0.3 km s$^{-1}$&\cite{kn98}\\
\hline
\end{tabular}
\end{adjustbox}	
\label{tab:properties}
\end{table*}

In ISM, phosphine (PH$_{3}$) is a relatively stable oblate symmetric top molecule. The rotational levels of PH$_{3}$ are given by two quantum numbers ($J$, $K$), and radiative transitions are allowed within the levels of the $K$ ladder ($\Delta$$J$ = 1, $\Delta$$K$ = 0). The electric dipole moment of PH$_{3}$ is 0.573 Debye \citep{de71}. \citet{sou20} claimed that PH$_{3}$ acts as a biosignature in the space. Except for our planet, evidence of PH$_{3}$ is also found in the atmospheres of Saturn and Jupiter with mixing ratios of 2 ppm and 0.6 ppm using the Voyager data \citep{mac05}. Subsequently, \citet{fle09} demonstrated the global distribution of PH$_{3}$ in the atmosphere of Saturn and Jupiter by using Cassini/CIRS observations. Recently, \cite{gap24} also found evidence of PH$_{3}$ in the atmosphere of Jupiter using Herschel/PACS. At high temperatures and pressures, PH$_{3}$ is produced in the deep atmosphere on large gas planets \citep{bre75, tar92}. Recently, \citet{gre21} reported the identification of the absorption line of PH$_{3}$ at a frequency of 266.944 GHz using ALMA and JCMT on the deck of Venus with a mixing ratio of 20 ppb. Several questions exist regarding the detection of PH$_{3}$ and the chemical models of the atmosphere of Venus. First, \citet{gre21} could not explain the formation of highly abundant PH$_{3}$ in the atmosphere of Venus by using steady-state and photochemical models. \citet{gre21} showed different abiotic chemical routes to explain the high abundance of PH$_{3}$ in the atmosphere of Venus. Subsequently, \citet{vil21} and several other authors raised questions regarding the spectroscopic data analysis by \citet{gre21}. \citet{vil21} and other authors clearly showed there is no evidance of PH$_{3}$ in the atmosphere of Venus. Subsequently, \cite{cor22} also attempted to search for the emission lines of other transitions of PH$_{3}$ at frequencies of 533 GHz and 1067 GHz using the NASA SOFIA aircraft, but they could not detect any absorption lines of PH$_{3}$. Therefore, \cite{cor22} estimated that the upper limit of PH$_{3}$ in Venus in the altitude range of 75--110 km is $\leq$0.8 ppb. Earlier, \citet{ol21} also searched for evidence of PH$_{3}$ towards the atmosphere of Mars, but they did not detect PH$_{3}$. The upper limit abundance of PH$_{3}$ towards Mars is $\leq$0.6 ppbv \citep{ol21}. Except for our solar system, the emission line of PH$_{3}$ was tentatively detected in the envelope of the carbon-rich star IRC+10216 using the IRAM 30 m single-dish telescope \citep{ag08}. This detection is tentative because \citet{ag08} did not properly identify the emission line of PH$_{3}$ at 266.944 GHz, owing to the limitation of the spectral resolution ($\sim$1.25 MHz) of IRAM. \citet{ag08} also claimed that IRC+10216 is one of the sources in the ISM where a very high abundance of PH$_{3}$ is still present because the PH$_{3}$/HCP abundance ratio is similar to the NH$_{3}$/HCN in the envelope of this source.

IRC+10216 (alternatively CW Leonis) is known as a carbon-rich asymptotic giant branch (AGB) star located at a distance of 130 pc \citep{men12}. This carbon-rich star is near the end of its evolution and is very close to being converted into a protoplanetary nebula \citep{sk98, os00}. The physical properties of IRC+10216 are presented in Table~\ref{tab:properties}. IRC+10216 loses mass at a very high rate (2$\times$10$^{-5}$ \(\textup{M}_\odot\)yr$^{-1}$) because the object is very close to the end of its AGB lifetime and this source is covered by an extensive circumstellar envelope (CSE) \citep{cr97}. This source is ideal for studying complex organic molecular lines because of its carbon-rich environment. Approximately 70 individual molecules have been detected in IRC+10216, including MgNC, AlNC, NaCN, AlCl, NaCl, KCl, AlF, and carbon-chain molecules \citep{cer87, cer00,zir02}.

In this letter, we present the first confirmation of interstellar phosphine (PH$_{3}$) towards the carbon-rich star IRC+10216 using ALMA. For the detection of the emission line of PH$_{3}$, we used the local thermodynamic equilibrium model (LTE). The observations and data reduction are presented in Section~\ref{obs}. The result of the detection of the rotational emission line of PH3 is shown in Section~\ref{res}. The discussion and conclusion are shown in Section~\ref{dis} and ~\ref{con}.

\begin{figure*}
\centering
\includegraphics[width=1.0\textwidth]{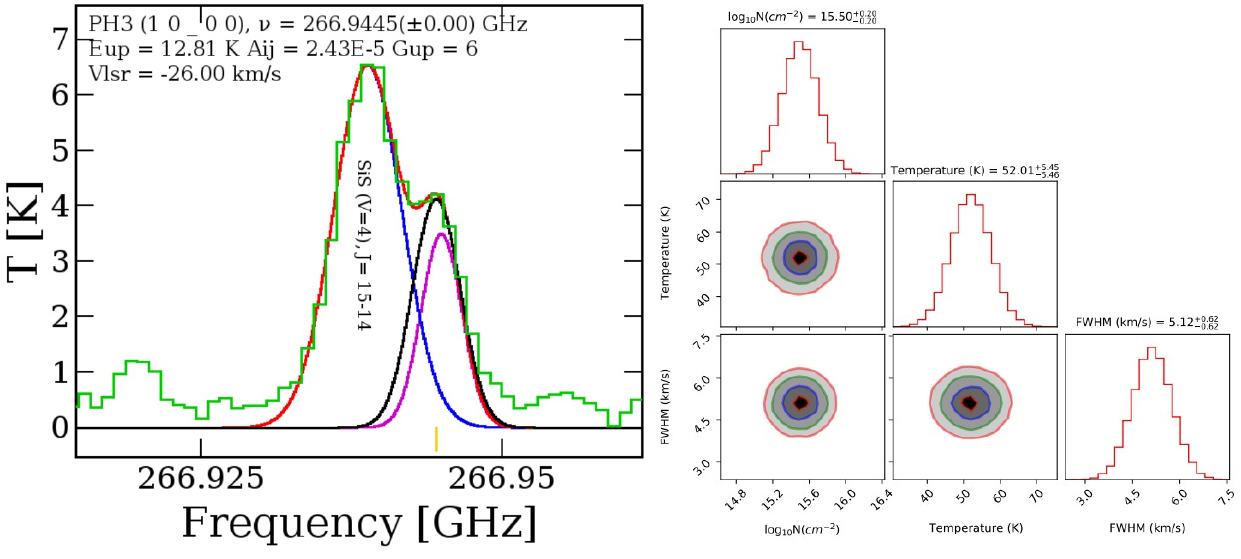}
\caption{Rotational emission line of PH$_{3}$ with transition $J$ = 1$_{0}$--0$_{0}$ towards IRC+10216 (left panel). The green lines represent the observed molecular spectrum of IRC+10216. The black spectra indicate the best-fit LTE model spectra for PH$_{3}$. The red spectrum is the global Gaussian model. The blue spectra are the Gaussian model corresponding to the SiS ($V$ = 4) emission line, and the violet spectra are the Gaussian model corresponding to the PH$_{3}$ emission line. The yellow vertical line indicates the rest frequency position of the PH$_{3}$. The right panel image shows the corner diagram plot based on MCMC fittings. Corner plots showing the 3D posterior probability distributions of column density in cm$^{-2}$, excitation temperature in K, and FWHM in km s$^{-1}$.}

\label{fig:ltespec}
\end{figure*} 

\section{Observations and data reductions}
\label{obs}
We used the cycle 0 archival data of IRC+10216, which was observed using a high-resolution ALMA in band 6 (PI: Cernicharo, Jose, ID: 2011.0.00229.S) with a 12 m array. This observation was performed on April 8, 2012, to study the emission lines of HCN with transitions $J$ = 3--2 and $J$ = 8--7 and observation of the dust formation zone in IRC+10216 with an on-source integration time of 32.76 min. The phase centre of IRC+10216 observation was ($\alpha,\delta$)$_{\rm J2000}$ = 09:47:57.406, +13:16:43.561. At the time of observation, 3C 279 and 3C 273 were used as bandpass and flux calibrators, respectively. J0854+201 was used as the phase calibrator. During the observation period, 16 antennas were used, with a minimum baseline of 15.7 m and a maximum baseline of 384.1 m. Observations were performed in the frequency ranges of 265.05--266.92 GHz, 266.15--268.03 GHz, 267.54--269.42 GHz, and 268.05--269.92 GHz with a spectral resolution of 976.56 kHz.

For data analysis, we used the Common Astronomy Software Application ({CASA 5.4.1}) with the ALMA data analysis pipeline \citep{mc07}. We applied the Perley-Butler 2017 flux calibration model for each baseline for flux calibration using the {\tt SETJY} task \citep{per17}. We also used the pipeline tasks HIFA\_bandpassflag and HIFA\_flagdata for flagging bad antenna data and channels, which were performed after flux and bandpass calibration. After a preliminary reduction of the data, we split the target data (IRC+10216) using the {\tt MSTRANSFORM} task with all the rest frequencies. We also used the {\tt UVCONTSUB} task to subtract continuum emission from the UV plane of the calibrated data. After data reduction, continuum emission images of IRC+10216 were created using the {\tt TCLEAN} task with a {\tt HOGBOM} deconvolver for line-free channels. Earlier, \citet{au15} discussed the dust continuum emission of IRC+10216 using the same data. Therefore, we do not discuss dust continuum emissions in this study. We also created spectral data cubes using the {\tt TCLEAN} task with the {\tt SPECMODE} = {\tt CUBE} parameter and Briggs weighting with a robust value of 0.5. We also applied a multiple self-calibration method to improve the RMS of the data cubes. To correct the primary beam pattern, we used the {\tt IMPBCOR} task.

\begin{figure}
	\centering  
	\includegraphics[width=0.48\textwidth]{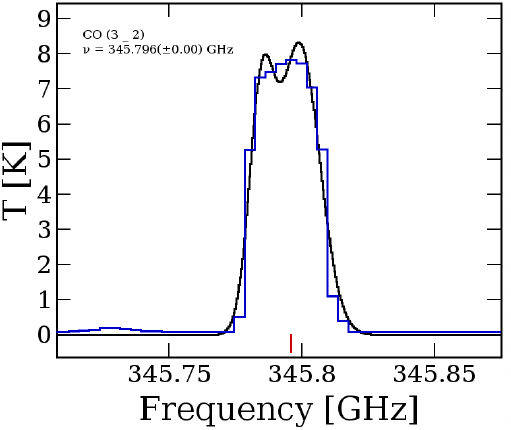}
	\caption{Rotational emission line of carbon monoxide (CO) with transition $J$ = 3--2 towards IRC+10216. The blue line indicates the observed spectra of CO, and the black spectrum is the best-fitting Gaussian model.}
	\label{fig:carbon}
\end{figure}

\section{Result}
\label{res}

\subsection{Identification of PH$_{3}$ towards IRC+10216}
To study the rotational emission line of PH$_{3}$, we only focus on the spectral data cube that was observed between the frequency range of 266.15 GHz and 268.03 GHz because the rest frequency of PH$_{3}$ ($J$ = 1$_{0}$--0$_{0}$) is 266.944 GHz. We extracted the molecular spectra of IRC+10216 from the spectral data cube to create a 2.3$^{\prime\prime}$ diameter circular region over the line-emitting region of the source, which is larger than the synthesized beam size of the spectral data cube. The synthesized beam size of the spectral data cube is 0.90$^{\prime\prime}$$\times$0.48$^{\prime\prime}$. The systemic velocity ($V_{\rm LSR}$) of IRC+10216 is --26.5 km s$^{-1}$ \citep{au15}. To identify the rotational emission line of PH$_{3}$, we used the local thermodynamic equilibrium (LTE) model spectra with the Cologne Database for Molecular Spectroscopy (CDMS) database \citep{mu05}. To fit the LTE spectra to the observed spectra of PH$_{3}$, we used the Markov chain Monte Carlo (MCMC) algorithm in CASSIS \citep{vas15}. The MCMC analysis using the CASSIS is well described in \citet{man24}. The MCMC method initializes by randomly selecting a seed point ($X_0$) in the three-dimensional parameter space. Then, it randomly chooses a nearby point ($X_1$) based on a variable step size, which is recalculated for each iteration. The $\chi^{2}$ value of the new state ($X_1$) is calculated, and if the ratio $p$ = $\chi^{2}$($X_0$)/$\chi^{2}$($X_1$) $>$ 1, the new state is accepted. However, even if $p$ $<$ 1, the new state may still be accepted with a certain probability. If the new state is rejected, the original state ($X_0$) remains, and another nearby point ($X_1$) is randomly selected. By allowing a finite probability of accepting a worse $\chi^{2}$ value, the algorithm avoids converging directly to a local minimum and instead ensures a more thorough exploration of the entire parameter space. During our MCMC analysis, we employed 1000 walkers, uniformly distributed within the specified parameter ranges, and ran the chains for a burning sequence of 20,000 steps to ensure convergence. The MCMC approach allows us to vary all parameters, including column density, excitation temperature, and full width at half maximum (FWHM) until the best fit is obtained. After the LTE spectral analysis, we detected the emission line of PH$_{3}$ at a frequency of 266.944 GHz with transition $J$ = 1$_{0}$--0$_{0}$ in the spectra of IRC+10216. We also observed that the emission line of PH$_{3}$ is closely associated with the emission line of SiS $V$ = 4 with transition $J$ = 15--14. The rest frequency of the emission line of SiS $V$ = 4 is 266.941 GHz, which was obtained from \citet{ag08}. Previously, \citet{ag08} and \citet{ag14} attempted to search the emission line of PH$_{3}$ ($J$ = 1$_{0}$--0$_{0}$) using the IRAM 30 m telescope, but the authors did not detect the proper peak of the emission line of PH$_{3}$ because of the limitation in the resolution of the IRAM. Our detection of the emission line of PH$_{3}$ using ALMA is the first confirmation of the presence of PH$_{3}$ in IRC+10216. After spectral analysis, we see that the emission line of PH$_{3}$ is non-blended. The upper state energy ($E_{u}$) and Einstain coefficients ($A_{ij}$) of the identified PH$_{3}$ transition are 12.81 K and 2.43$\times$10$^{-5}$ s$^{-1}$. The full-width half maximum (FWHM) of the LTE spectra is 5.12$\pm$0.62 km s$^{-1}$. According to the LTE model, the best-fit column density of PH$_{3}$ is (3.15$\pm$0.20)$\times$10$^{15}$ cm$^{-2}$ with an excitation temperature of 52$\pm$5 K and a source size of 0.90$^{\prime\prime}$. Our estimated excitation temperature of PH$_{3}$ is similar to the excitation temperature of another P-bearing molecule, CP, which was estimated by \citet{mil08}. The excitation temperature of PH$_{3}$ is relatively low because the AGB star is near the end of its evolution. The peak and integrated intensities of the detected emission line of PH$_{3}$ are 4.26$\pm$0.12 K and 20.46$\pm$0.56 K km s$^{-1}$ respectively. The optical depth of the spectra of PH$_{3}$ is 5.81$\times$10$^{-2}$. The estimated optical depth indicates that the identified rotational emission line of PH$_{3}$ is optically thin. The LTE-fitted rotational emission line of PH$_{3}$ towards IRC+10216 is shown in Figure~\ref{fig:ltespec}. In addition, we created a corner plot corresponding to the LTE fitting values. The corner plot shows the 3D posterior probability distributions of column density in cm$^{-2}$, excitation temperature in K, and FWHM in km s$^{-1}$ of PH$_{3}$ towards IRC+10216, which is shown in the right panel of Figure~\ref{fig:ltespec}.

\begin{figure}
	\centering
	\includegraphics[width=0.5\textwidth]{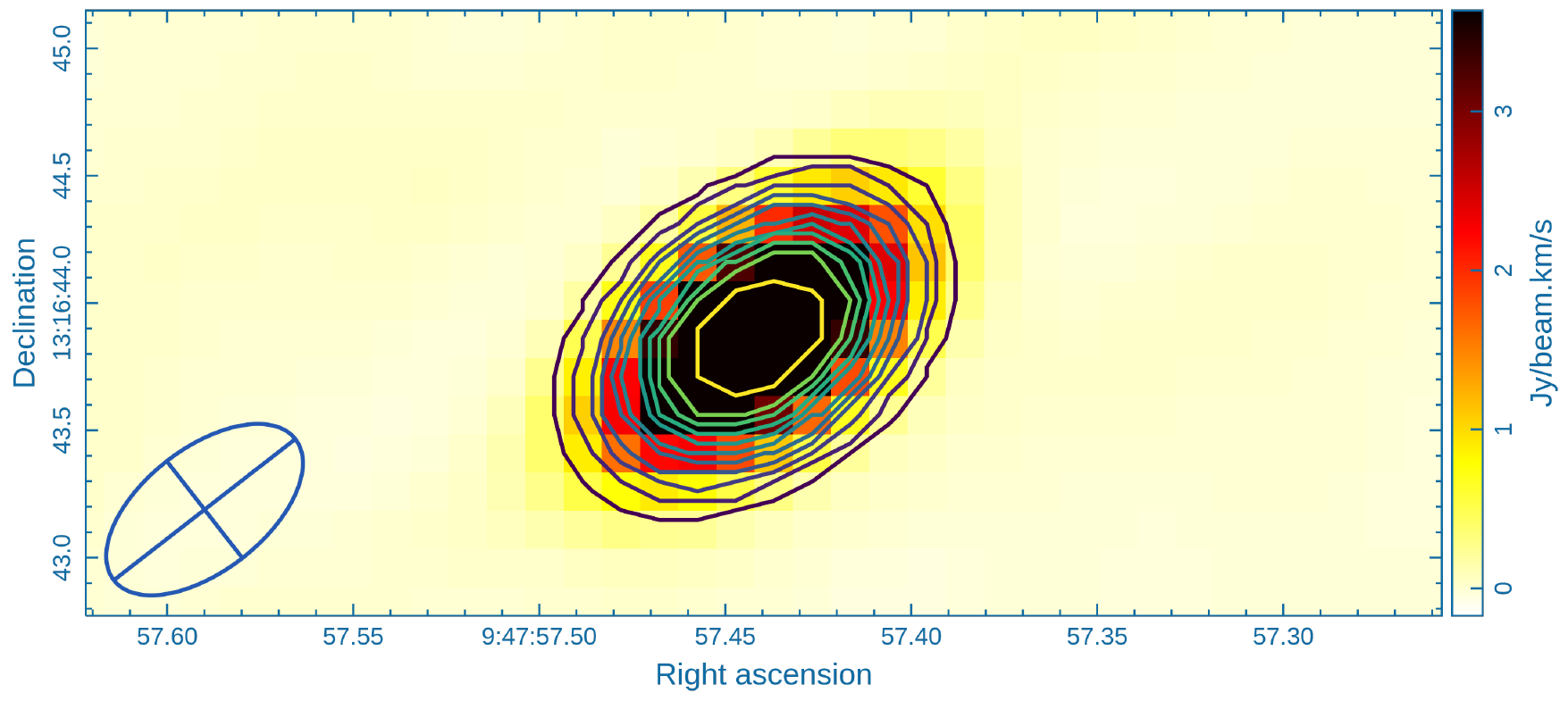}
	\caption{Integrated emission map of PH$_{3}$ towards IRC+10216. The contour levels are started at 3$\sigma$ and are increased by a factor of $\surd$2. The blue circle represents the synthesized beam of the integrated emission map.}
	\label{fig:intmap}
\end{figure}

\subsection{Estimation of the column density of molecular H$_{2}$ towards IRC+10216}
To determine the column density of molecular H$_{2}$ towards IRC+10216, we used the following equation: 
\begin{equation}
{\rm N(H_2)=2.0 \times 10^{20} \times \frac{{\rm W(^{12}CO)}}{K~kms^{-1}}},
\label{nh2}
\captionsetup{labelformat=empty}
\end{equation}
where W($^{12}$CO) is the integrated intensity of $^{12}$CO ($J$ = 3--2) at the corresponding velocity intervals. This equation was taken from \citet{is21}. To determine the emission line properties of CO, we used the 2016.1.00251.S (PI: Vlemmings Wouter) ALMA data. The analysis of this ALMA data is well described in \citet{si22}. The emission line of CO towards IRC+10216 is shown in Figure~\ref{fig:carbon}. We fitted a Gaussian model over the emission line of CO and estimated that the integrated intensity of the CO line is 190.39 K kms$^{-1}$. Using the above equation, the estimated column density of H$_{2}$ towards IRC+10216 is (3.80$\pm$0.58)$\times$10$^{22}$ cm$^{-2}$.

\subsection{Abundance of PH$_{3}$ towards IRC+10216}
To derive the fractional abundance of PH$_{3}$, we used the column density of PH$_{3}$ inside the 0.90$^{\prime\prime}$ synthesized beam, which was divided by the column density of H$_{2}$. The fractional abundance of PH$_{3}$ with respect to molecular H$_{2}$ towards IRC+10216 is (8.29$\pm$1.37)$\times$10$^{-8}$, where the column density of molecular H$_{2}$ towards IRC+10216 is (3.80$\pm$0.58)$\times$10$^{22}$ cm$^{-2}$. Previously, \citet{ag08} and \citet{ag14} estimated the tentative abundance of PH$_{3}$ towards IRC+10216, which varied between $\sim$10$^{-9}$ and $\sim$10$^{-8}$. Our estimated fractional abundance of PH$_{3}$ using ALMA data is similar to those of \citet{ag14} but one order of magnitude higher than those of \citet{ag08}. Previously, \cite{lef16} attempted to detect the emission line of PH$_{3}$ at a frequency of 266.944 GHz using the IRAM 30 m telescope towards the solar-type star-forming region L1157, however, they could not successfully detect the emission line of PH$_{3}$. The upper limit of the abundance of PH$_{3}$ towards L1157 is $\leq$10$^{-9}$ \citep{lef16}. Recently, \citet{fu24} also searched the emission line of PH$_{3}$ ($J$ = 1$_{0}$--0$_{0}$) towards L1544 using the ALMA, but they could not detect PH$_{3}$. The upper limit column density and abundance of PH$_{3}$ towards L1544 are $\leq$7.6$\times$10$^{10}$ cm$^{-2}$ and $\leq$6.7$\times$10$^{-12}$, respectively \citep{fu24}. Therefore, we confirm that IRC+10216 is the only source in the ISM where evidence of PH$_{3}$ is found.
	
\subsection{Spatial distribution of PH$_{3}$}
We created an integrated emission map of PH$_{3}$ ($J$ = 1$_{0}$--0$_{0}$) towards IRC+10216 using the {\tt IMMOMENTS} task. In the {\tt IMMOMENTS} task, we used the channel ranges of the spectral data cubes, where the emission lines of PH$_{3}$ were identified. The integrated emission map of PH$_{3}$ is shown in Figure~\ref{fig:intmap}. The emission map clearly shows that the emission line of PH$_{3}$ originated from the inner envelope of IRC+10216. We also fitted a 2D Gaussian over the integrated emission map of PH$_{3}$ using the {\tt IMFIT} task. The following equation is used to estimate the emitting region of PH$_{3}$\\
\begin{equation} 
\theta_{S}=\sqrt{\theta^2_{50}-\theta^2_{\text{beam}}} 
\end{equation}
In the above equation, $\theta_{50} = 2\sqrt{A/\pi}$ indicates the diameter of the circle whose area ($A$) surrounds the $50\%$ line peak of PH$_{3}$ and $\theta_{\text{beam}}$ indicates the half-power width of the synthesized beam of the integrated emission map of PH$_{3}$ \citep{man23}. The size of the emitting region of PH$_{3}$ is 0.83$^{\prime\prime}$. The synthesized beam size of the integrated emission map of PH$_{3}$ is 0.90$^{\prime\prime}$$\times$0.48$^{\prime\prime}$. We observed that the emitting region of PH$_{3}$ is slightly smaller than the synthesized beam size of the integrated emission map. This indicates that the detected emission line of PH$_{3}$ is not spatially resolved towards IRC+10216. Therefore, we cannot draw any conclusions regarding the spatial distribution morphology of PH$_{3}$ towards IRC+10216. Higher angular and spatial resolution observations with better spectral resolution are required to understand the chemical morphology of PH$_{3}$ towards IRC+10216.

\section{Discussion}
\label{dis}
\subsection{Possible formation mechanism of PH$_{3}$}
Previous chemical modelling studies have indicated that PH$_{3}$ is an important biomolecule that plays a major role in the production of other P-bearing molecules using the surface chemistry of the interstellar grains \citep{cha94, ao12, nu20}. Earlier, \citet{nu20} showed that PH$_{3}$ is created on the grain surface rather than in the gas phase. \citet{cha94} and \citet{nu20} proposed a possible formation mechanism for PH$_{3}$ via the hydrogenation of atomic P on the grain surface of the highly dense part of star-formation regions and hot molecular cores via the following reactions. \\\\
P + H $\longrightarrow$PH, $\Delta$ = --277 KJ mol$^{-1}$~~~~~~~~~~~~~~~ (1)\\\\
PH + H $\longrightarrow$ PH$_{2}$, $\Delta$ = --347 KJ mol$^{-1}$~~~~~~~~~~~(2)\\\\
PH$_{2}$ + H $\longrightarrow$ PH$_{3}$, $\Delta$ = --326 KJ mol$^{-1}$~~~~~~~~~(3)\\\\    
Since reactions 1--3 are barrierless, PH$_{3}$ can be produced on the grain surface. At low temperatures ($\sim$10 K), the hydrogen atom (H) can diffuse and react with other compounds on the grain surface \citep{ha13}. Previously, no P-bearing molecules, including PH$_{3}$, were observed in the solid state. \citet{tur15, tur18, tur19} experimentally observed that solid PH$_{3}$ is converted into phosphoric acid, diphosphate, and methyl phosphonic acid at low temperatures. \citet{cha20} also showed that the PH$_{3}$ is destroyed due to the reaction of C$^{+}$ (C$^{+}$ + PH$_{3}$ $\longrightarrow$PH$_{3}$$^{+}$ + C) and H$^{+}$ (H$^{+}$ + PH$_{3}$ $\longrightarrow$PH$_{3}$$^{+}$ + H). Earlier, \citet{ag14} computed the chemical model of PH$_{3}$ in the IRC +10216 environment using reaction 3, and they found that the model abundance of PH$_{3}$ is 1.0$\times$10$^{-8}$. Our observed abundance of PH$_{3}$ towards IRC+10216 using the ALMA is very close to the modelled abundance of PH$_{3}$, which was estimated by \citet{ag14}. This indicates PH$_{3}$ is formed via hydrogenation of PH$_{2}$ on the grain surface of IRC+10216.

\begin{figure}
	\centering
	\includegraphics[width=0.5\textwidth]{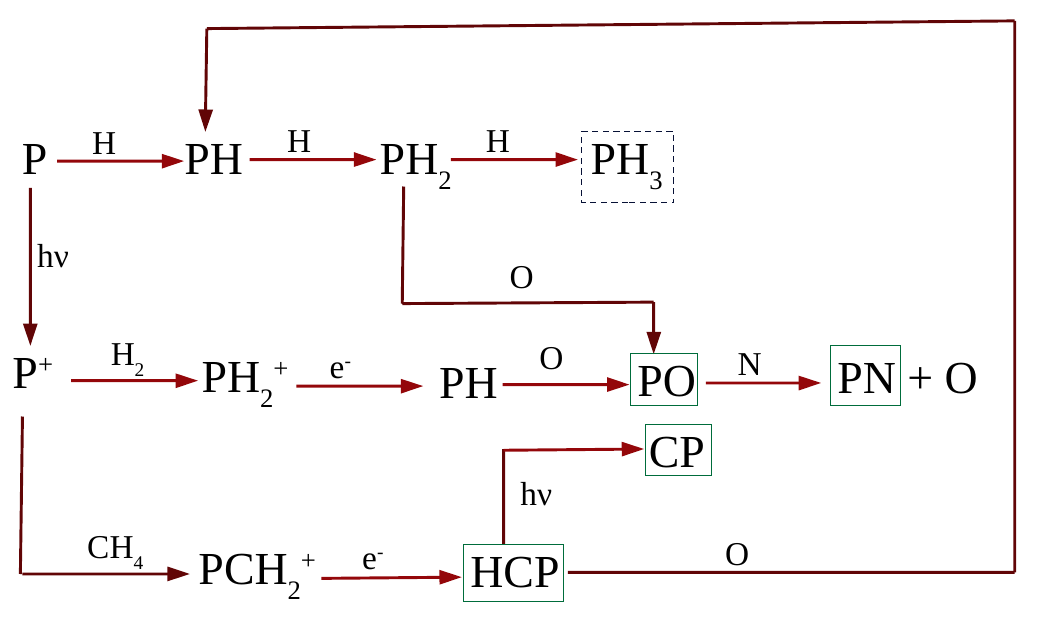}
	\caption{The proposed gas phase and grain surface chemical network for the formation of PH$_{3}$ and link with other P-bearing molecules. The green box indicates P-bearing molecules which were previously detected in IRC+10216.}
	\label{fig:chemicalnetwork}
\end{figure}

\subsection{Chemical link-up between PH$_{3}$ and other molecules}
Previously the emission lines of HCP, CP, PO, and PN were detected towards the IRC+10216 \citep{mat87, gu90, mil08}. We created a chemical network to understand the prebiotic chemistry of PH$_{3}$ and the chemical link-up between all detected P-bearing species towards IRC+10216, as shown in Figure~\ref{fig:chemicalnetwork}. The chemical reactions are taken from \citet{cha94}, \citet{ao12}, \citet{ha13}, \citet{nu20}, and UMIST 2012 \citep{mc13} molecular reaction database. Our chemical network clearly shows that PH acts as a possible precursor of PH$_{3}$, PN, and PO. Previously, \citet{ag14} showed that the gas phase chemistry is not sufficient for the formation of PH$_{3}$. Therefore, there is a high chance of the formation of PH$_{3}$ via hydrogenation of PH$_{2}$ on the grain surface of IRC+10216. A new chemical model using the gas-grain chemistry and quantum chemical studies using the density functional theory (DFT) is needed with proper formation and destruction pathways of PH$_{3}$ with reaction rates to understand the chemical evolution and proper formation and destruction pathways of PH$_{3}$ towards IRC+10216.
	
\section{Summary and conclusion}
\label{con}
In this letter, we present the first confirmation of the rotational emission line of PH$_{3}$ towards the carbon-rich AGB star IRC+10216 at a frequency of 266.944 GHz using the ALMA band 6. The abundance of PH$_{3}$ towards IRC+10216 is (8.29$\pm$1.37)$\times$10$^{-8}$. We discuss the possible formation pathways of PH$_{3}$ and we claim that PH$_{3}$ may be formed via the hydrogenation of PH$_{2}$ on the grain surface of IRC+10216. The confirmed detection of PH$_{3}$ indicates that the grain surface chemistry is sufficient for the production of other P-bearing molecules because PH$_{3}$ and PH act as possible precursors of other P-bearing molecules. Our chemical network shows that PH$_{3}$, PN, and PO are chemically connected to the PH. A detailed spectral line study and chemical modelling are required to understand other P-bearing molecules towards IRC+10216, which will be carried out in our next follow-up study.

\section*{Acknowledgement}
We thank the anonymous referee for the helpful comments that improved the manuscript. A.M. acknowledges the Swami Vivekananda Merit cum Means Scholarship, Government of West Bengal, India, for financial support for this research. The emission spectra of PH$_{3}$ are available on our \href{https://github.com/astrochemistry/PH3-in-IRC-10216-using-ALMA}{github} repository. The emission map and chemical network of PH$_{3}$ within this paper is available from the corresponding author upon reasonable request. This paper makes use of the following ALMA data: ADS /JAO.ALMA\#2011.0.00229.S and 2016.1.00251.S. ALMA is a partnership of ESO, NSF (USA), and NINS (Japan), together with NRC (Canada), MOST and ASIAA (Taiwan), and KASI (Republic of Korea), in cooperation with the Republic of Chile. The JAO is operated by ESO, AUI/NRAO, and NAOJ.

\bibliographystyle{aasjournal}

\begin{thebibliography}{}
	\expandafter\ifx\csname natexlab\endcsname\relax\def\natexlab#1{#1}\fi
	\providecommand{\url}[1]{\href{#1}{#1}}
	\providecommand{\dodoi}[1]{doi:~\href{http://doi.org/#1}{\nolinkurl{#1}}}
	\providecommand{\doeprint}[1]{\href{http://ascl.net/#1}{\nolinkurl{http://ascl.net/#1}}}
	\providecommand{\doarXiv}[1]{\href{https://arxiv.org/abs/#1}{\nolinkurl{https://arxiv.org/abs/#1}}}
	
\bibitem[\protect\citeauthoryear{Altwegg et al.}{2016}]{al16}Altwegg, K., Balsiger, H., Bar-Nun, A., et al. 2016, \href{https://www.science.org/doi/10.1126/sciadv.1600285}{SciA}, 2, e1600285	

\bibitem[\protect\citeauthoryear{Ag{\'u}ndez et al.}{2008}]{ag08}Ag{\'u}ndez, M., Cernicharo, J., Pardo, J.~R., Gu{\'e}lin, M., Phillips, T.~G., 2008, \href{https://doi.org/10.1051/0004-6361:200810193}{A\&A}, 485, L33-L36

\bibitem[\protect\citeauthoryear{Ag{\'u}ndez et al.}{2014}]{ag14}Ag{\'u}ndez, M., Cernicharo, J., Decin, L., Encrenaz, P., Teyssier, D., 2014, \href{https://iopscience.iop.org/article/10.1088/2041-8205/790/2/L27/meta}{ApJL}, 790, 2, L27

\bibitem[\protect\citeauthoryear{Aota \& Aikawa}{2012}]{ao12}Aota, T., \& Aikawa, Y. 2012, \href{https://iopscience.iop.org/article/10.1088/0004-637X/761/1/74/meta}{ApJ}, 761, 74

\bibitem[\protect\citeauthoryear{Ag{\'u}ndez et al.}{2015}]{au15}Ag{\'u}ndez, M., Cernicharo, J., Quintana-Lacaci, G., Velilla Prieto, L., Castro-Carrizo, A., Marcelino, N., Gu{\'e}lin, M. 2015, \href{https://iopscience.iop.org/article/10.1088/0004-637X/814/2/143/meta}{ApJ}, 814, 143

\bibitem[\protect\citeauthoryear{Bregman et al.}{1975}]{bre75}Bregman, J., Lester, D., \& Rank, D. 1975, \href{https://adsabs.harvard.edu/full/1975ApJ...202L..55B}{ApJL}, 202, L55

\bibitem[\protect\citeauthoryear{Crosas \& Menten}{1997}]{cr97}Crosas, M., Menten, K., 1997, \href{https://iopscience.iop.org/article/10.1086/304256/meta}{ApJ}, 483, 913

\bibitem[\protect\citeauthoryear{Cordiner et al.}{2022}]{cor22}Cordiner, M. A., Villanueva, G. L., Wiesemeyer, H., et al., 2022, \href{https://doi.org/10.1029/2022GL101055}{GRL}, 49, 22

\bibitem[\protect\citeauthoryear{Charnley \& Millar}{1994}]{cha94}Charnley, S. B., \& Millar, T. J. 1994, \href{https://doi.org/10.1093/mnras/270.3.570}{MNRAS}, 270, 570

\bibitem[\protect\citeauthoryear{Chantzos et al.}{2020}]{cha20}Chantzos, J., Rivilla, V. M., Vasyunin, A., et al. 2020, \href{https://doi.org/10.1051/0004-6361/201936531}{A\&A}, 633, A54

\bibitem[\protect\citeauthoryear{Cernicharo et al.}{2000}]{cer00}Cernicharo, J., Gu$\acute{e}$lin, M., \& Kahane, C. 2000, \href{https://doi.org/10.1051/aas:2000147}{A\&AS}, 142, 181

\bibitem[\protect\citeauthoryear{Cernicharo \& Guelin}{1987}]{cer87}Cernicharo, J., \& Guelin, M. 1987, \href{https://adsabs.harvard.edu/full/1987A\&A...183L..10C}{A\&A}, 183, L10

\bibitem[\protect\citeauthoryear{De Beck et al.}{2013}]{de13}De Beck, E., Kami{\'n}ski, T., Patel, N. A., et al. 2013, \href{https://doi.org/10.1051/0004-6361/201321349}{A\&A}, 558, A132

\bibitem[\protect\citeauthoryear{Davies et al.}{1971}]{de71}Davies, P. B., Neumann, R. M., Wofsy, S. C., \& Klemperer, W. 1971, \href{https://doi.org/10.1063/1.1676614}{J. Chem. Phys}, 55, 3564

\bibitem[\protect\citeauthoryear{Fontani et al.}{2016}]{fon16}Fontani, F., Rivilla, V. M., Caselli, P., Vasyunin, A., \& Palau, A. 2016, \href{https://iopscience.iop.org/article/10.3847/2041-8205/822/2/L30/meta}{ApJL}, 822, L30

\bibitem[\protect\citeauthoryear{Fletcher et al.}{2009}]{fle09}Fletcher, L. N., Orton, G. S., Teanby, N. A., \& Irwin, P. G. J. 2009, \href{https://doi.org/10.1016/j.icarus.2009.03.023}{Icarus}, 202, 543

\bibitem[\protect\citeauthoryear{Furuya \& Shimonishi}{2024}]{fu24} Furuya, K., \& Shimonishi, T., 2024, \href{https://iopscience.iop.org/article/10.3847/2041-8213/ad50cc/meta}{ApJL}, 968, L19

\bibitem[\protect\citeauthoryear{Gapp et al.}{2024}]{gap24} Gapp, C., Rengel, M., Hartogh, P., Sagawa, H., Feuchtgruber, H., Lellouch, E., Villanueva, G.~L., 2024, \href{https://doi.org/10.1051/0004-6361/202347345}{A\&A}, 688, A10

\bibitem[\protect\citeauthoryear{Groenewegen et al.}{2012}]{gro12}Groenewegen M. A. T. et al., 2012, \href{https://doi.org/10.1051/0004-6361/201219604}{A\&A}, 543, L8

\bibitem[\protect\citeauthoryear{Gu{\'e}lin et al.}{1995}]{gue95}Gu{\'e}lin M., Forestini M., Valiron P., Ziurys L. M., Anderson M. A., Cernicharo J., Kahane C., 1995, \href{https://adsabs.harvard.edu/full/1995A\%26A...297..183G}{A\&A}, 297, 183

\bibitem[\protect\citeauthoryear{Gu{\'e}lin et al.}{1990}]{gu90}Gu{\'e}lin, M., Cernicharo, J., Paubert, G., \& Turner, B. E. 1990, \href{https://adsabs.harvard.edu/full/1990A\%26A...230L...9G}{A\&A}, 230, L9-L11

\bibitem[\protect\citeauthoryear{Greaves et al.}{2021}]{gre21}Greaves, J. S., Richards, A. M. S., Bains, W., et al. 2021, \href{https://doi.org/10.1038/s41550-020-1174-4}{NatAs}, 5, 655

\bibitem[\protect\citeauthoryear{Halfen et al.}{2008}]{hal08}Halfen, D. T., Clouthier, D. J., \& Ziurys, L. M. 2008, \href{https://iopscience.iop.org/article/10.1086/588024/meta}{ApJ}, 677, L101

\bibitem[\protect\citeauthoryear{Hama \& Watanabe}{2013}]{ha13}Hama, T., \& Watanabe, N. 2013, \href{https://doi.org/10.1021/cr4000978}{ChRv}, 113, 8783

\bibitem[\protect\citeauthoryear{Isequilla et al.}{2021}]{is21}Isequilla, N.~L., Ortega, M.~E., Areal, M.~B., Paron, S. 2021, \href{https://doi.org/10.1051/0004-6361/202039974}{A\&A}, 649, A139

\bibitem[\protect\citeauthoryear{Knapp et al.}{1998}]{kn98}Knapp, G. R., Young, K., Lee, E., Jorissen, A., 1998, \href{https://iopscience.iop.org/article/10.1086/313111/meta}{ApJS}, 117, 209

\bibitem[\protect\citeauthoryear{Lefloch et al.}{2016}]{lef16}Lefloch, B., Vastel, C., Viti, S., et al. 2016, \href{https://doi.org/10.1093/mnras/stw1918}{MNRAS}, 462, 3937

\bibitem[\protect\citeauthoryear{Ladjal et al.}{2010}]{la10}Ladjal, D. et al., 2010, \href{https://doi.org/10.1051/0004-6361/201014658}{A\&A}, 518, L141

\bibitem[\protect\citeauthoryear{Menten et al.}{2012}]{men12}Menten K., Reid M. J., Kaminski T., Claussen M. J., 2012, \href{https://doi.org/10.1051/0004-6361/201219422}{A\&A}, 543, A73

\bibitem[\protect\citeauthoryear{Maci{\'a} et al.}{1997}]{mac97}Maci{\'a}, E., Hernández, M., \& Or{\'o}, J. 1997, \href{https://doi.org/10.1023/A:1006523226472}{OLEB}, 27, 459

\bibitem[\protect\citeauthoryear{Milam et al.}{2008}]{mil08}Milam, S. N., Halfen, D. T., Tenenbaum, E. D., et al. 2008, \href{https://iopscience.iop.org/article/10.1086/589135/meta}{ApJ}, 684, 618


\bibitem[\protect\citeauthoryear{Maci{\'a} et al.}{2005}]{mac05}Maci{\'a}, E. 2005, \href{https://doi.org/10.1039/B416855K}{Chem. Soc. Rev.}, 34, 691

\bibitem[\protect\citeauthoryear{McMullin et al.}{2007}]{mc07}McMullin, J. P., Waters, B., Schiebel, D., Young, W., \& Golap, K. 2007, in Astronomical Society of the Pacific Conference Series, Vol. 376, \href{https://ui.adsabs.harvard.edu/abs/2007ASPC..376..127M/abstract}{Astronomical Data Analysis Software and Systems XVI}, ed. R. A. Shaw, F. Hill, \& D. J. Bell, 127

\bibitem[\protect\citeauthoryear{M\"uller et al.}{2005}]{mu05} M\"uller, H. S. P., SchlM$\ddot{o}$der, F., Stutzki, J. Winnewisser, G., 2005, \href{https://doi.org/10.1016/j.molstruc.2005.01.027}{J. Mol. Struct}, 742, 215

\bibitem[\protect\citeauthoryear{Manna \& Pal}{2024}]{man24}Manna, A., \& Pal, S., 2024,  \href{https://doi.org/10.1007/s12036-023-09989-x}{J. Astrophys. Astron}, 45, 3

\bibitem[\protect\citeauthoryear{Manna et al.}{2023}]{man23} Manna, A., Pal, S., Viti, S., Sinha, S., 2023, \href{https://doi.org/10.1093/mnras/stad2185}{MNRAS}, 525, 2229


\bibitem[\protect\citeauthoryear{Matthews et al.}{1987}]{mat87}Matthews H. E., Feldman P. A., Bernath P. F., 1987, \href{https://adsabs.harvard.edu/full/record/seri/ApJ../0312/1987ApJ...312..358M.html}{ApJ}, 312, 358

\bibitem[\protect\citeauthoryear{McElroy et al.}{2013}]{mc13} McElroy, D., Walsh, C., Markwick, A.~J., Cordiner, M.~A., Smith K., Millar, T.~J., 2013, \href{https://doi.org/10.1051/0004-6361/201220465}{A\&A}, 550, A36

\bibitem[\protect\citeauthoryear{Nguyen et al.}{2020}]{nu20}Nguyen, T., Oba, Y., Shimonishi, T., Kouchi, A., Watanabe, N. 2020, \href{https://iopscience.iop.org/article/10.3847/2041-8213/aba695/meta}{ApJL}, 898, L52

\bibitem[\protect\citeauthoryear{Osterbart et al.}{2000}]{os00}Osterbart R., Balega Y. Y., Bl$\ddot{o}$cker T., Men’shchikov A. B., Weigelt G., 2000, \href{https://articles.adsabs.harvard.edu/pdf/2000A%26A...357..169O}{A\&A}, 357, 169

\bibitem[\protect\citeauthoryear{Olsen et al.}{2021}]{ol21} Olsen, K.~S., Trokhimovskiy, A., Braude, A.~S., Korablev, O.~I., Fedorova, A.~A., Wilson, C.~F., Patel, M.~R., et al., 2021, \href{https://doi.org/10.1051/0004-6361/202140868}{A\&A}, 649, L1

\bibitem[\protect\citeauthoryear{Perley \& Butler}{2017}]{per17}Perley, R. A., Butler, B. J. 2017, \href{https://iopscience.iop.org/article/10.3847/1538-4365/aa6df9/meta}{ApJs}, 230, 1538

\bibitem[\protect\citeauthoryear{Rivilla et al.}{2020}]{riv20}Rivilla, V. M., Drozdovskaya, M. N., Altwegg, K., et al. 2020, \href{https://doi.org/10.1093/mnras/stz3336}{MNRAS}, 492, 1180

\bibitem[\protect\citeauthoryear{Siebert et al.}{2022}]{si22}Siebert, M~A., Van de Sande, M., Millar, T~J., Remijan, A~J. 2022, \href{https://iopscience.iop.org/article/10.3847/1538-4357/ac9e52/meta}{ApJ}, 941, 90

\bibitem[\protect\citeauthoryear{Skinner et al.}{1998}]{sk98}Skinner C. J., Meixner M., Bobrowsky M., 1998, \href{https://doi.org/10.1046/j.1365-8711.1998.300004l29.x}{MNRAS}, 300, 29

\bibitem[\protect\citeauthoryear{Sousa-Silva et al.}{2020}]{sou20}Sousa-Silva, C., Seager, S., Ranjan, S., et al. 2020, \href{https://doi.org/10.1089/ast.2018.1954}{AsBio}, 20, 235

\bibitem[\protect\citeauthoryear{Turner et al.}{1990}]{tur90}Turner, B. E., Tsuji, T., Bally, J., Guelin, M., \& Cernicharo, J. 1990, \href{https://adsabs.harvard.edu/full/1990ApJ...365..569T}{ApJ}, 365, 569

\bibitem[\protect\citeauthoryear{Tenenbaum et al.}{2007}]{ten07}Tenenbaum, E. D., Woolf, N. J., \& Ziurys, L. M. 2007, \href{https://iopscience.iop.org/article/10.1086/521361}{ApJ}, 666, L29

\bibitem[\protect\citeauthoryear{Tenenbaum et al.}{2008}]{ten08}Tenenbaum, E. D., \& Ziurys, L. M. 2008, \href{https://iopscience.iop.org/article/10.1086/589973}{ApJ}, 680, L121

\bibitem[\protect\citeauthoryear{Turner et al.}{2019}]{tur19}Turner, A. M., Abplanalp, M. J., Bergantini, A., et al. 2019, \href{https://www.science.org/doi/10.1126/sciadv.aaw4307}{SciA}, 5, eaaw4307

\bibitem[\protect\citeauthoryear{Turner et al.}{2015}]{tur15}Turner, A. M., Abplanalp, M. J., Chen, S. Y., et al. 2015, \href{https://doi.org/10.1039/C5CP02835C}{PCCP}, 17, 27281

\bibitem[\protect\citeauthoryear{Turner et al.}{2018}]{tur18}Turner, A. M., Bergantini, A., Abplanalp, M. J., et al. 2018, \href{https://doi.org/10.1038/s41467-018-06415-7}{Nat Commun}, 9, 3851

\bibitem[\protect\citeauthoryear{Tarrago et al.}{1992}]{tar92}Tarrago, G., Lacome, N., L{\'e}vy, A., et al. 1992, \href{https://ui.adsabs.harvard.edu/abs/1992JMoSp.154...30T/abstract}{JMoSp}, 154, 30

\bibitem[\protect\citeauthoryear{Villanueva et al.}{2021}]{vil21}Villanueva, G.L., Cordiner, M., Irwin, P.G.J. et al. 2021, \href{https://www.nature.com/articles/s41550-021-01422-z}{NatAs}, 5, 631--635

\bibitem[\protect\citeauthoryear{Vastel et al.}{2015}]{vas15}Vastel, C., Bottinelli, S., Caux, E., Glorian, J. -M., Boiziot, M., 2015, \href{https://ui.adsabs.harvard.edu/abs/2015sf2a.conf..313V/abstract}{Proceedings of the Annual meeting of the French Society of Astronomy and Astrophysics}, 313-316

\bibitem[\protect\citeauthoryear{Ziurys et al.}{2002}]{zir02}Ziurys, L. M., Savage, C., Highberger, J. L., et al. 2002, \href{https://iopscience.iop.org/article/10.1086/338775/meta}{ApJ}, 564, L45

\bibitem[\protect\citeauthoryear{Ziurys et al.}{2018}]{zi18}Ziurys, L. M., Schmidt, D. R., \& Bernal, J. J. 2018, \href{https://iopscience.iop.org/article/10.3847/1538-4357/aaafc6/meta}{ApJ}, 856, 169


\end{thebibliography}

\end{document}